\documentstyle[aps]{revtex}
\newcommand{\ds}{\displaystyle}
\newcommand{\dsf}{\ds\frac}

\newcommand{\beq}{\begin{equation}}
\newcommand{\eeq}{\end{equation}}

\large

\begin{document}
\large
\begin{center}
\Large\bf
Nonlinear Thermomagnetic Waves in Type II Superconductors
\vskip 0.1cm
{\normalsize\bf N.A.\,Taylanov}\\
\vskip 0.1cm
{\large\em Theoretical Physics Department and
Institute of Applied Physics,\\
National University of Uzbekistan,\\
E-mail: taylanov@iaph.silk.org}
\end{center}
\begin{center}
\bf €bstract
\end{center}

\begin{center}
\mbox{\parbox{14cm}{\small

        This work is devoted to analysis of the nonlinear dynamics
of interrelated thermal and electromagnetic perturbations in a
superconductor caused by dissipative effects involved in the viscous
motion of the magnetic flux. The structure and evolution of the stationary
thermomagnetic waves propagating in the superconductor is studied based
on a system of nonlinear equations describing dynamics of the instability
development in the system studied. The apperance of these structures is
qualititively described and the wave propagation velocity is estimated.
The affect a thermoactivated flux creep to the structures of the nonlinear
thermomagnetic wave propagation is studied. It is shown that taking into
account a nonlinear dependency of the current density $j$ on the electric
field strength $E$ would not qualitatively change the main results, since
the character of the equilibrium state on the phase plane remains esentially
the same.
}}
\end{center}
\vskip 0.5cm

At present superconducting systems with high critical field strengs and
current densites are widely implemented in various advenced technologies.
Hovever, successful operation of the superconducting materials is only
possible provided that special measures are taken to prevent a system
from the thermal or magnetic breakage of superconductivity and transition
to a normal (resistive)state [1].
For this reason, one of the main problems in the investigation of
properties of superconductors is that of predicting the superconducting
state breacage caused by dissipative and nonlinear effects related to
viscous motions of the magnetic flux.
This explains a considerable interest in the study of dissipative and
nonlinear effects in superconductors that has apose in recent years.

        This work is devoted to analysis of the nonlinear dynamics
of interrelated thermal and electromagnetic perturbations in a
superconductor caused by dissipative effects involved in the viscous
motion of the magnetic flux.
The structure and evolution of the stationary thermomagnetic waves propagating
in the superconductor is studied based on a system of nonlinear equations
describing dynamics of the instability development in the system is studied.

        The evolution of thermal and electromagnetic perturbations in a
superconductor is described by a set involving the thermal conductivity
equation [2-5]

\beq
\nu\dsf{dT}{dt}=\nabla(\kappa\nabla T)+\vec j\vec E\,,
\eeq
the Maxwell equation
\beq
\dsf{4\pi}{c^2}\dsf{d\vec j}{dt}=\Delta \vec E\,,
\eeq
and the related equation of the critical state
\beq
\vec j=\vec j_{á}(T,\vec H)+\vec j_{r}(\vec E)\,.
\eeq
Here $\nu=\nu(T)$ is the heat capacity, $\kappa=\kappa(T)$ is the thermal conductivity
coefficient, $\vec j_{c}$ is the critical current density, and $\vec j_{r}$
is the current density in the resistive state.

        The above system is essentially nonlinear because the right-hand
part of Eq.(1) contains a term describing the Joule heat evolution
in the region of a resistive phase.
Such a set (1)-(3) of nonlinear parabolic differential equations in
partial derivatives has no exact analytical solution.

        Let us consider a planar semi-infinite sample $(x>0)$ placed in an
external magnetic field $\vec H=(0, 0, H_{e})$ growing at a constant rate
$\dsf{d\vec H}{dt}=const$. According to Maxwel's equation (2), there is a vortex
electric field $\vec E=(0, E_e, 0)$ in the sample, directed parallel to the
current density $\vec j$: $\vec E\parallel \vec j$; where $H_e$ is the
amplitude of the external magnetic field and $E_e$ is the amplitude of the
external electric field.

      For the automodel propagating waves of the type $\xi(x,t)= x-vt$
the initial differential Eqs. (1)-(3) acquire the following form

\beq
-\nu v\frac{dT}{d\xi}=\frac{d}{d\xi}\left[\kappa\frac{dT}{d\xi}\right]+jE\,,
\eeq
\beq
\frac {dE}{d\xi}=-\frac{4\pi v}{c^2}j\,,
\eeq
\beq
E=\frac{v}{c}H\,.
\eeq
       A set of boundary conditions corresponding to Eqs. (4)-(6) is as
follows:

\beq
\begin{array}{l}
T(\xi\rightarrow+\infty)=T_0\,,\qquad  \dsf{dT}{d\xi}(\xi\rightarrow-\infty)=0\,,\\
\quad\\
E(\xi\rightarrow+\infty)=0\,,\qquad   E(\xi\rightarrow-\infty)=E_e\,,
\end{array}
\eeq

          Let us use the Bean-London model of the critical state for the
dependence $j_{c}(T,H)$; $\dsf{dj_c}{dH}=0$ [6]. According to this model
the dependence $j_{c}(T)$ can be represented as
$j_{c}(T)=j_0[1-a(T-T_{0})]$: where $j_{0}$ is the equilibtium current density,
$a$ is the constant parameter and $T_0$ is the temperature of the cooling
medium.
A characteristic field dependence of $j_r(E)$ in the region of sufficiently
strong electric fields $(E>E_f)$ can be aproximated by a piecewise
linear function $j_r\approx\sigma_f E$, where $\sigma_f$ is the effictive
conductivity. In the field of small field strengths $(E<E_f)$ we assume a
relationship $j_{r}(E)\approx j_1\ln\dsf{E}{E_0}$,to be valid with $j_1$ being a
characteristic local current density scatter (related to a pinning force
inhomogeneity) on the order of  $j_1\approx 0,01j_c$ and $E_0$ being a
constant. This relationship between $j_{r}$ and $E$ is due to a thermoactivated
early experiments (see,e.g.[1]). The thermoactivated flux creep is principal
mechanism of the energy dissipation during the magnetic flux penetration
deep into the sample. This factor must affect the character of the
nonlinear thermomagnetic wave propagation in the system studied.

     Excluding the variables $T(\xi)$ and $H(\xi)$, using Eqs.(4) and (6) ,
and taking into account the boundary conditions (7), we obtain an equation
describing the electric field $E(\xi)$ distribution:
\beq
\frac{d^2E}{dz^2}+\beta\left[1+\frac{j_1}{\sigma_d E}\tau
\right]\frac{dE}{dz}+\beta^2\tau
\left[\frac{j_0}{\sigma_d}+\frac{j_1}{\sigma_d}\ln\frac{E}{E_0}
\right]=\frac{E^2}{2E_\kappa}\,.
\eeq
The dimensionless variables used in Eq. (8) are defined as follows:
$$
z=\frac{\xi}{L}\,,\quad
\beta=\frac{vt_\kappa}{L}\,,\quad
t_\kappa=\frac{\nu L^2}{\kappa}\,,\quad
\tau=\frac{4\pi\sigma_d\kappa}{c^2\nu}\,,\quad
E_\kappa=\frac{\kappa}{aL^2}\,,\quad
L=\frac{cH_e}{4\pi j_0}\,,
$$
where $L$ is the depth of the magnetic flux penetration into the sample,
$\sigma_{d}$ is the differential conductivity.
        The corresponding equation of state is obtained using a relationship
\beq
\Omega E^2=X(E)=1+\dsf{j_1}{j_0}\ln\dsf{E}{E_0},
\eeq
where
$\Omega=\dsf{\sigma_d}{2\beta^2\tau j_0E_\kappa}$.

As seen from the plots of $X(E)$ versus $E$ (see figure), there exists a
single point of intersection of the curves $y=\Omega E^2$ and
$y=1+(j_1/j_0)\ln(E/E_0)$, which corresponds to a single stable equilibrium
state. The stability of this state is determined by the sign of derivative
$d^2E/dz^2$ in the vicinity of the equilibrium point.
The wave velocity $v_E$ can be determined from stationary $(\dsf{dE}{dz}=0)$
Eq.(8) with an allowance for the boundary conditions [7]:
\beq
v_E=\frac{L}{t_\kappa} E_e\left[2\tau\frac{j_0E_\kappa}{\sigma_d}\left(1+
\frac{j_1}{j_0}\ln\frac{E_e}{E_0}\right)\right]^{-1/2}\,.
\eeq

         Now we can use Eq. (6) and readily derive an equation describing
the field distribution for a nonlinear $H$ - wave. The wave velocity $v_H$
of this $H$- wave is given by the formula

\beq
v_H=\frac{cE}{H_e}
\exp\left[
\frac{\sigma_d}{2\tau j_1}E_\kappa\left(\frac{LH_e}{ct_\kappa}\right)^2
-\frac{j_0}{j_1}\right]\,.
\eeq
As seen, the wave velocity $v_H$ exponentially increases with the field
amplitude $H_e$. For a suffuciently small amplitude $H_e<H_a$, where
\beq
H_a=\frac{ct_\kappa}{L}\left[\frac{2\tau j_0E_\kappa}{\sigma_d}\right]^{1/2}
\eeq
the wave velocity is negigible small, which corresponds to the case of
the thermoactivated flux creep.
For $H_e=H_a$, the wave propagates at a finite constant velocity. In this
case, the maximum heating of a superconductor in the region immiadiately
in front of the wave is described by the relationship

\beq
\frac{T-T_0}{T_0}=\frac{H_e^{2}}{8\pi\nu T_0}\,.
\eeq

        In conluding, we must note that taking into account a nonlinear
dependency of the current density $j$ on the electric field strength $E$
would not qualitatively change the main results, since the character
of the equilibrium state on the phase plane remains esentially the same
(see [2]).

\end{document}